# Dual Modulation Faraday Rotation Spectroscopy of HO$_2$

Brian Brumfield[1], Wenting Sun[2], Yiguang Ju[2], and Gerard Wysocki[1*]

[1]*Department of Electrical Engineering, Princeton University, Princeton NJ 08544, USA*
[2]*Mechanical and Aerospace Engineering Department, Princeton University, Princeton NJ, USA*
*Corresponding Author: gwysocki@princeton.edu

The technique of dual modulation Faraday rotation spectroscopy has been applied for near shot-noise limited detection of HO$_2$ at the exit of an atmospheric pressure flow reactor. This was achieved by combining direct current modulation at 51 kHz of an external cavity quantum cascade laser system with 610 Hz modulation of the magnetic field generated by a Helmholtz coil. The DM-FRS measurement had a 1.8 times better signal-to-noise ratio than an AC-FRS measurement acquired under identical flow reactor conditions. Harmonic detection of the FRS signal also eliminated the substantial DC-offset associated with electromagnetic intereference pick-up from the Helmholtz coils that is observed in the AC-FRS spectrum. A noise equivalent angle of $4\times10^{-9}$ rad Hz$^{-1/2}$ was observed for the DM-FRS measurement, and this corresponds to a 3σ detection limit of 0.2 ppmv Hz$^{-1/2}$.

The development of sensitive optical diagnostics for measurement of hydroperoxyl radical (HO$_2$) in combustion processes is important for the validation of chemical kinetic models that describe the oxidation of fuels in the low-temperature regime [1]. Unfortunately, HO$_2$ remains a challenging chemical species to directly quantify in combustion processes using conventional absorption spectroscopic methods [2-4].

We have recently quantified the HO$_2$ concentration at the exit of an atmospheric pressure flow reactor by using mid-IR Faraday rotation spectroscopy (FRS) [5]. FRS is a laser-based magneto-optical technique [5-9] that relies on the introduction of magnetic circular birefringence (MCB) in the vicinity of optical transitions belonging to radicals in a gas-phase sample. In conventional FRS an AC magnetic field is utilized (AC-FRS), and the signal from the modulated sample MCB is demodulated at the magnetic field frequency by using phase sensitive detection (PSD). Additional improvement in the AC-FRS detection limit is hindered by 1/f noise due to the low modulation frequency of the electromagnetic coils. Many AC-FRS systems also suffer from electromagnetic interference pick-up (EMI) in the detection electronics. This can generate a significant background on the FRS signal, and long-term drift in the background can impair the long-term sensitivity of the FRS measurement.

To overcome the technical limitations of AC-FRS, a new detection method has been developed that relies on high-frequency wavelength modulation of the laser ($f_L$) with simultaneous modulation of the magnetic field ($f_M$) [10-12]. In this dual modulation (DM-)FRS method the FRS signal at $f_M$ is encoded by amplitude modulation on the carrier frequency $f_L$. Demodulation at $f_L$ using PSD permits retrieval of the AM signal at $f_M$, and the AM signal can then be demodulated at $f_M$ using PSD. Because $f_L \gg f_M$, the 1/f laser intensity noise at $f_L$ is less than at $f_M$, and the EMI pick-up is also eliminated by demodulation at $f_L$. The effectiveness of DM-FRS compared to AC-FRS has already been realized by near shot-noise limited measurements of NO in a medical breath analysis application [11]. In this Letter we report the successful application of DM-FRS to detection of HO$_2$ from an atmospheric flow reactor.

The experimental layout used for the collection of DM-FRS spectra is shown in Fig. 1. Overall this experimental layout is similar to that presented in Ref. [5] for AC-FRS measurements. A Q-branch spectral feature in the ν$_2$ vibrational band of $HO_2$ is targeted at 1396.90 cm$^{-1}$ using IR radiation from an EC-QCL (Daylight Solutions, model 21074-MHF). Slow tuning of the laser frequency is achieved by step-scanning the voltage applied to a piezoelectric transducer that controls the angle of an intra-cavity diffraction grating. High-frequency modulation of the laser wavelength at 51 kHz ($f_L$) is achieved via modulation of the laser current by applying a voltage sine wave from a function generator (Tektronix AFG3102) to an external current modulation input on the laser housing. The output radiation for the laser passes through a pair of Brewster plate polarizers (InnoPho PGC-5) to clean-up the linear polarization of the laser. The light is then passed between the openings in a Helmholtz coil arrangement that provides a 105 Gauss RMS field at the exit of an atmospheric flow reactor that is modulated at a frequency of 610 Hz. The laser beam passes within 1.7 mm of the reactor exit and then exits the measurement region through an opening in the electromagnetic coil. Two wire grid polarizers were used as a single polarization analyzer, and provide a minimum measured extinction of 30,000:1. The light transmitted by the polarization analyzer is focused onto a photodiode (Vigo PVI-4TE-8).

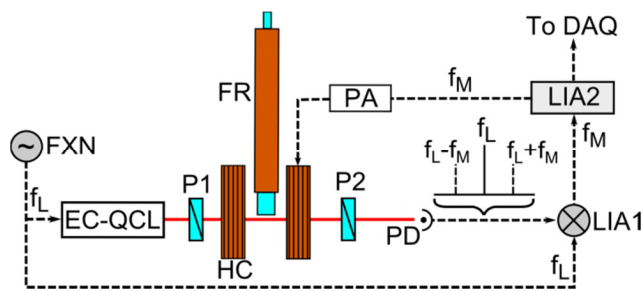

Fig. 1. Experimental layout used for collection of DM-FRS spectra of HO$_2$. The labels in the figure describe the following: FXN) function generator, EC-QCL) external-cavity quantum cascade laser, P1&2) polarizers 1 and 2, HC) Helmholtz coils, FR) flow reactor, PD) photodiode, PA) power amplifier, LIA1&2) lock-in amplifier 1 and 2, DAQ) digital acquisition, $f_L$) laser wavelength modulation frequency, $f_M$) magnetic field modulation frequency.

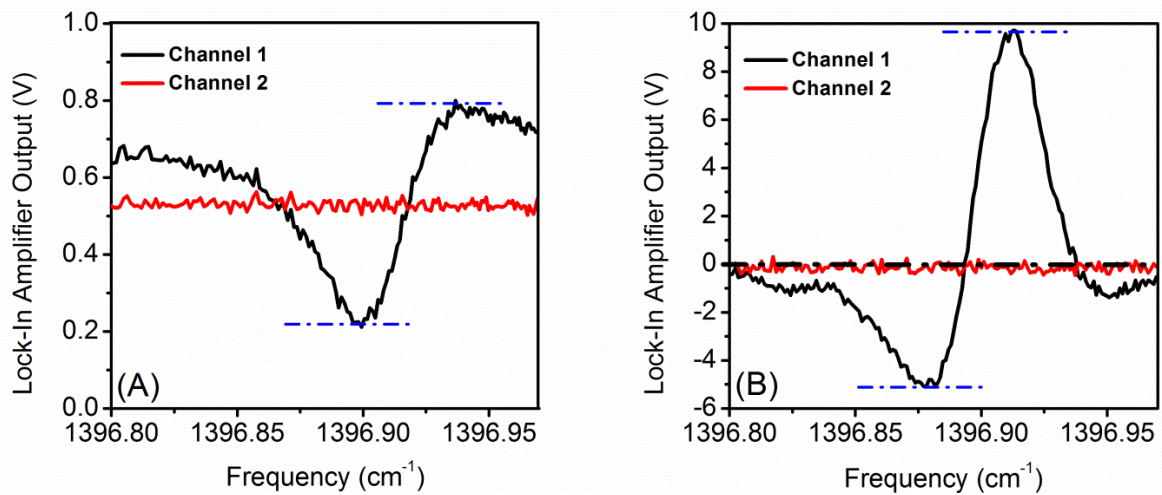

Fig. 2. $HO_2$ spectra collected using AC-FRS (Fig. 2A) or the DM-FRS (Fig. 2B) technique. Both spectra were collected at a flow reactor temperature of 600 K and a gas mixture of 0.5% DME, 9.5% $O_2$, and 90% He. The ENBW set by the second lock-in amplifier time constant was 1.25 Hz, and the voltages for the in-phase (Channel 1) and quadrature (Channel 2) components were measured using the analog outputs of the instrument. The uncrossing angles for the polarization analyzer were 0.5° and 1° for Fig. 2A and Fig. 2B respectively. In both figures the two output channels from the second lock-in amplifier are shown. Channel 2 (shown in red) corresponds to the background channel with no signal. The peak-to-peak signal value was determined in both spectra by using the minimum and maximum values shown in blue dashed horizontal lines.

The voltage signal from the photodetector is sent to the input of a lock-in amplifier (Signal Recovery Model 7280) for demodulation at $f_L$. This lock-in amplifier is used as a frequency mixer. A time constant of 100 μs was used to ensure that the measurement bandwidth of the lock-in would transmit without attenuation the AM sidebands generated at $f_L \pm f_M$ to a second lock-in amplifier (Signal Recovery Model 7265). This second lock-in amplifier demodulated the output signal from the first lock-in amplifier at $f_M$, and an output voltage from the in-phase and quadrature components of the lock-in were measured using the GPIB interface or analog output of the lock-in amplifier. The second lock-in amplifier also provided the reference sine wave sent to a power amplifier that was used to the drive the resonant LC circuit producing the modulated magnetic field in the experiment.

Figure 2 shows an example of FRS spectra collected with the experimental layout shown in Fig. 1. The DM-FRS spectrum shown in Fig. 2B was collected first and then followed by collection of the AC-FRS spectrum shown in Fig. 2A. The flow reactor conditions were the same for both spectra so that the same $HO_2$ sample concentration was measured. However, the uncrossing angle for the polarization analyzer (P2) went from 1° to 0.5° moving from the DM-FRS to AC-FRS measurement due to the increased laser intensity noise. This was necessary to meet the optimum signal-to-noise (SNR) ratio requirement in FRS where the analyzer is uncrossed until the laser intensity noise contribution is equal to the noise contribution from the photodetector [7]. The SNR for the DM-FRS and AC-FRS spectra was 99:1 and 55:1 respectively. The SNR achieved using DM-FRS is 1.8 times better than what was achieved using AC-FRS. From Fig. 2 it is also apparent that the large DC offset observed in Fig. 2A is almost entirely eliminated in Fig. 2B. The remaining offset observed in Fig. 2B is an offset voltage associated with the analog output and not the demodulated signal. This has been verified by acquiring DM-FRS spectra and recording the demodulated signals through the GPIB interface of the second lock-in amplifier.

The observed noise equivalent angle for the AC-FRS measurement ($\Theta_{AC-FRS,NEA}$) is estimated to be $1.4 \times 10^{-8}$ rad Hz$^{-1/2}$, and this is 3.5 times greater than the DM-FRS NEA of $4 \times 10^{-9}$ rad Hz$^{-1/2}$. The smaller laser intensity noise value reflects a reduction in 1/f noise associated with the laser intensity by shifting the measurement frequency from 610 Hz to 51 kHz. The shot-noise limited NEA ($\Theta_{SNEA}$) given the effective optical power transmitted through the system is $9 \times 10^{-10}$ rad Hz$^{-1/2}$. Therefore the DM-FRS measurement is only 4.4 times from the fundamental limit of detection.

Non-linear fitting of the AC-FRS signal in Fig. 2A leads to an estimate of an HO$_2$ concentration of 8 ppmv. Using this concentration estimate, the estimated 3σ detection limit for the DM-FRS spectrum shown in Fig. 2B is 0.2 ppmv Hz$^{-1/2}$. The DM-FRS detection limit is comparable to the detection limits reported in the previous AC-FRS measurement [5], however, the DM-FRS measurement was performed with a third of the optical path and 18% lower magnetic field strength compared to Ref. [5]. This indicates that the DM-FRS measurement is at least 3 times as sensitive as the previously reported AC-FRS measurement.

While the SNR advantage of DM-FRS represents a marginal improvement, more significant is the elimination of the EMI offset in the signal. For the AC-FRS signal shown in Fig. 2A, the DC offset in the FRS signal is equal to the EMI pick-up. Due to its relative magnitude with respect to the signal, drift in the offset can have a significant impact on applications where single-point measurements at the peak of the spectral feature are performed. While appropriate shielding of the detection electronics can help mitigate the DC-offset, DM-FRS offers a way to do this without shielding while also offering an improvement in the SNR. It is important to realize that if the EMI pick-up is sufficient to saturate the photodetector pre-amplifier it will still be necessary to first reduce the EMI pick-up through shielding and/or re-positioning of the pre-amplifier with respect to the electromagnet.

Despite the 3.5 times reduction in the laser intensity noise for the DM-FRS measurement, only a 1.8 times improvement is observed in the SNR in comparison to the AC-FRS measurement. This occurs because there are two parameters that can reduce the DM-FRS signal strength: the modulation amplitude of the laser wavelength and losses associated with harmonic detection. The dependence of the DM-FRS signal strength on the modulation amplitude is related to the half-width-at-half-maximum (HWHM) of the AC-FRS lineshape, and this is analogous to the relationship between the modulation amplitude and the HWHM of an absorption lineshape observed using wavelength modulation spectroscopy (WMS) [13]. For the DM-FRS spectrum in Fig. 2B, a voltage sine wave with a peak-to-peak voltage of 4 V was applied to the current modulation input of the EC-QCL. The relationship between the applied voltage sine wave and the wavelength modulation depth can be determined by calculating the first harmonic of the AC-FRS spectrum ($H_1(\tilde{\nu})$) in Fig. 2A using the following integral [14]:

$$H_1(\tilde{\nu}) = \left(\frac{2}{\pi}\right) \int_0^\pi G(\tilde{\nu} + \Delta\tilde{\nu} \cos\theta) \cos\theta \, d\theta \qquad (1)$$

where $G(\tilde{\nu})$ is the AC-FRS lineshape and $\Delta\tilde{\nu}$ is the modulation amplitude of the laser. In eqn. (1) the optical frequency is expressed in units of cm$^{-1}$. For a given value of $\Delta\tilde{\nu}$, the $H_1(\tilde{\nu})$ spectrum can be evaluated numerically using the experimental AC-FRS spectrum in Fig. 2A for $G(\tilde{\nu})$. Because the AC-FRS and DM-FRS spectra in Fig. 2 were acquired under identical flow reactor conditions and in less than 5 minutes, it is possible to scale $H_1(\tilde{\nu})$ to simulate the DM-FRS spectrum if differences in the electrical gain and optical powers between the two detection methods are taken into account. The expression that is used to scale $H_1(\tilde{\nu})$ from eqn. (1) is:

$$H_1(\text{DM-FRS}) = H_1(\tilde{\nu}) \times \frac{\theta_{DM-FRS}}{\theta_{AC-FRS}} \times \frac{G_{LIA1}}{2^{1/2}} \times T. \qquad (2)$$

where θ$_{AC-FRS}$ and θ$_{DM-FRS}$ represent the uncrossing angle of the polarization analyzer in the AC-FRS and DM-FRS experimental configuration respectively, G$_{LIA1}$ is the gain for LIA1, and T represents the additional transmission losses of the AM signal associated with LIA1. The ratio of the uncrossing angle of the polarization analyzer is necessary because the FRS signal scales linearly with optical power in the small angle limit. The value of $H_1(\tilde{\nu})$ is divided by $2^{1/2}$ because the output of LIA1 is the RMS value of the signal demodulated at $f_L$. The transmission factor T is 0.98, and was determined empirically by passing a well-defined AM signal from a function generator through the cascaded lock-in amplifier arrangement used for DM-FRS measurements.

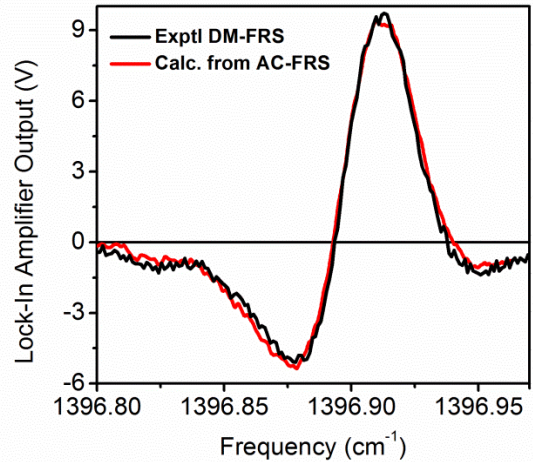

Fig. 3. Comparison between the experimental DM-FRS spectrum (black trace, Exptl. DM-FRS) from Fig. 2B and a calculated DM-FRS spectrum (red trace, Calc. from AC-FRS) obtained by numerical evaluation of the AC-FRS spectrum in Fig. 2A using eqns. (1) and (2) with a $\Delta \tilde{\nu}$ value of 0.016 cm$^{-1}$.

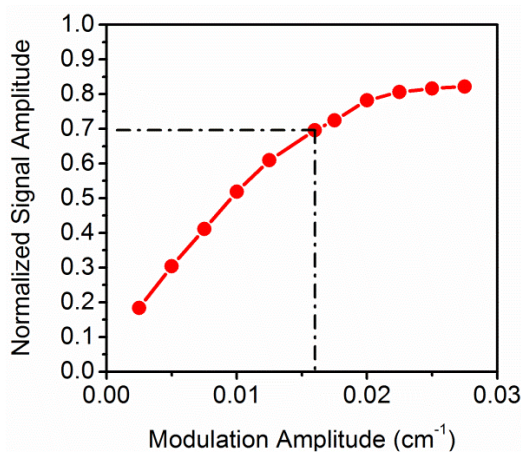

Fig. 4. Normalized peak-to-peak signal amplitude for the calculated DM-FRS signal plotted against the modulation amplitude of the laser. The calculated DM-FRS signal was normalized to the peak-to-peak AC-FRS signal in Fig. 2A. Dashed lines indicate the simulation data point corresponding to the estimated modulation amplitude of the laser when collecting the DM-FRS spectrum shown in Fig. 2B.

The only unknown parameter to convert the observed AC-FRS signal to the DM-FRS is $\Delta\tilde{v}$. A program was written in MatLab to numerically evaluate the result of eqn. (2) as $\Delta\tilde{v}$ was manually adjusted. A value of 0.016 cm$^{-1}$ (480 MHz) for $\Delta\tilde{v}$ generates a calculated DM-FRS spectrum that is in excellent agreement with the experimentally obtained DM-FRS spectrum. To determine the total signal loss due to 1f harmonic detection of the AC-FRS signal, the numerical calculation of the DM-FRS signal was carried out for multiple values of $\Delta\tilde{v}$. The results of the modeling are summarized in Fig. 4. At the estimated modulation amplitude of 0.016 cm$^{-1}$ the peak-to-peak signal amplitude for the DM-FRS spectrum is only 0.7 times that of the AC-FRS spectrum. By accounting for the signal loss due harmonic detection of the AC-FRS signal, the RMS output from LIA1, and transmission losses associated with LIA1, an estimated SNR improvement of 1.7 times for DM-FRS detection. This is in good agreement with the experimentally observed SNR improvement of of 1.8.

Fig. 4. shows that a maximum for the normalized signal occurs near $\Delta\tilde{v} = 0.028$ cm$^{-1}$, which is roughly twice the estimated HWHM for the AC-FRS spectrum in Fig. 2A. At a modulation amplitude of 0.028 cm$^{-1}$, the DM-FRS signal is 0.82 times that of the AC-FRS signal. It was not possible to reach this optimum modulation amplitude in the current work because the maximum peak-to-peak voltage that can be applied to the modulation input of the EC-QCL system. DM-FRS spectra were also collected at at an $f_L$ of 22.62 kHz and 100 kHz under identical flow reactor conditions to verify that no further optimization of the SNR was possible by avoiding roll-off in the modulation amplitude associated with larger values of $f_L$. The DM-FRS signal did not suffer from attenuation going from an $f_L$ of 22.62 kHz to 51 kHz, and only a 3% reduction in signal was observed going from 51 kHz to 100 kHz. The measurement noise is similar for at all three values of $f_L$. Given these results, it is clear that the modulation amplitude and value of $f_L$ are close to optimum for DM-FRS detection of HO$_2$ in roughly 1 atm of He using the EC-QCL system.

For measurements of HO$_2$ relevant to combustion chemistry it would be useful to carry out FRS measurements at pressures > 1 atm. The HWHM of the AC-FRS signal will increase because of additional pressure broadening under these conditions, and an EC-QCL system will be unable to target to the optimum modulation amplitude. In this situation it will be necessary to use a DFB-QCL in order to retain the SNR performance advantage of DM-FRS over AC-FRS detection. The higher obtainable modulation frequency of the diode laser may also lead to an additional SNR advantage by permitting shot-noise limited detection.

DM-FRS has been successfully used for near shot-noise limited measurements of HO$_2$ at the exit of an atmospheric flow reactor. With DM-FRS it is possible to reduce the HO$_2$ detection limit by a factor of 1.8 compared to AC-FRS. While this improvement is marginal, the large DC offset attributed to EMI pick-up in the detection electronics was nearly eliminated in the DM-FRS measurement. Single-point measurements using line-locking should permit DM-FRS HO$_2$ detection limits down into the 10 ppbv range within reasonable integration times (~100 seconds). Future application of DM-FRS to studies of HO$_2$ will also require the development of a modified numerical model of the AC-FRS signal that will account for nth harmonic detection of the FRS signal.

This work was jointly supported by the Princeton Environmental Institute through the Siebel Energy Challenge Award, the Andlinger Center for Energy and the Environment, and the DOE NETL University Turbine System Research grant.